\documentclass[
reprint,
superscriptaddress,
nofootinbib,
amsmath,amssymb,
aps,
]{revtex4-2}

\usepackage{graphicx}
\usepackage{hyperref}
\usepackage{dcolumn}
\usepackage{bm}
\usepackage[mathlines]{lineno}
\usepackage{verbatim}
\usepackage{hyperref}
\hypersetup{colorlinks,linkcolor={blue},citecolor={blue},urlcolor={blue}}
\usepackage{soul}
\sethlcolor{yellow} 
\usepackage{float}

\begin{document}
\preprint{APS/123-QED}
\title{The study of $0\nu\beta\beta$ decay of $^{136}$Xe using nonclosure approach in nuclear shell model}
\author{S. Sarkar}
\email{shahariar.ph@sric.iitr.ac.in}
\affiliation{Department of Physics, Indian Institute of Technology Roorkee, Roorkee - 247667, Uttarakhand, India}
\author{Y. Iwata}
\affiliation{Osaka University of Economics and Law, Yao, Osaka 581-0853, Japan}
\author{K. Jha}
\affiliation{Department of Physics, Medi-Caps University, Pigdamber, Rau, Indore - 453331, Madhya Pradesh, India}
\author{R. Chatterjee}
\affiliation{Department of Physics, Indian Institute of Technology Roorkee, Roorkee - 247667, Uttarakhand, India}
\date{\today}
\begin{abstract}
In this investigation, we compute the nuclear matrix elements (NMEs) relevant to the light neutrino-exchange mechanism governing neutrinoless double beta ($0\nu\beta\beta$) decay in $^{136}$Xe. Our method is based on the nonclosure approach within the interacting nuclear shell model framework. This approach considers the genuine effects arising from the excitation energies of two hundred states for each spin-parity of the intermediary nucleus $^{136}$Cs. All computations are performed using the effective shell model Hamiltonian GCN5082.
To understand the impact of nuclear structure on $0\nu\beta\beta$ decay, we explore the dependence of the NME on various factors, including the number of intermediate states and their spin-parity characteristics. We identify an optimal closure energy of approximately 3.7 MeV for the $0\nu\beta\beta$ decay of $^{136}$Xe that reproduces the nonclosure NME using the closure approach. The calculated total NME for the light neutrino-exchange $0\nu\beta\beta$ decay of $^{136}$Xe is 2.06 with the CD-Bonn short-range correlation (SRC).
These results can be valuable for future experimental investigations into the $0\nu\beta\beta$ decay of $^{136}$Xe.
\end{abstract}

\maketitle


\section{\label{sec:level1}Introduction}
The $0\nu\beta\beta$ decay is a rare type of weak nuclear decay that can occur in specific even-even nuclei such as $^{136}$Xe \cite{RevModPhys.95.025002,Dolinski:2019nrj,Vergados:2016hso,engel2017status}. In this process, two neutrons within the parent nucleus are simultaneously converted into two protons and two electrons, with no emission of neutrinos. The absence of neutrinos in the final state violates the law of lepton number conservation, making it a process of great interest to physicists. The detection of $0\nu\beta\beta$ decay in experiments would provide evidence for the Majorana nature of neutrinos. Majorana neutrinos, hypothesized as their own antineutrinos, could elucidate the unexpectedly small mass of neutrinos in various theoretical particle physics models. Moreover, exploring $0\nu\beta\beta$ decay could provide insights into the absolute scale of the effective Majorana neutrino mass, potentially revealing clues about the overall mass hierarchy of neutrinos. Currently, the most stringent upper limit on the effective Majorana neutrino mass ($\langle\rm m_{\beta\beta}\rangle$) is between 36 and 156 meV. This limit is derived from the $0\nu\beta\beta$ half-life of $^{136}\rm Xe$, $T_{1/2}^{0\nu} \ge 2.3\times 10^{26}$ years, using NMEs from various nuclear models, as reported by the KamLAND-Zen experiment \cite{PhysRevLett.130.051801}.

A number of decay mechanisms have been proposed for the $0\nu\beta\beta$ decay, such as standard light neutrino-exchange mechanism, supersymmetric particle exchange mechanisms, left-right symmetric mechanisms, etc. \cite{Vergados:2016hso}. In this study, we focus on the light neutrino-exchange mechanism. The decay rate for $0\nu\beta\beta$ decay mechanisms is related to NMEs and the absolute mass of the neutrino. These NMEs are typically calculated using theoretical nuclear many-body models \cite{engel2017status}. Some of the widely used models are the quasiparticle random phase approximation (QRPA) \cite{vsimkovic20130}, the interacting shell-model (ISM) \cite{PhysRevLett.100.052503,PhysRevC.93.024308,PhysRevLett.113.262501,PhysRevLett.116.112502}, the interacting boson model (IBM 2) \cite{PhysRevC.79.044301,PhysRevLett.109.042501}, the generator coordinate method (GCM) \cite{PhysRevLett.105.252503,PhysRevC.90.054309,PhysRevC.91.024316}, the energy density functional (EDF) theory \cite{PhysRevLett.105.252503,PhysRevC.90.054309}, the relativistic energy density functional (REDF) theory \cite{PhysRevC.90.054309,PhysRevC.91.024316}, the projected Hartree-Fock Bogolibov model (PHFB) \cite{PhysRevC.82.064310}, $ab$ $initio$ Quantum Monte Carlo (QMC) technique \cite{PhysRevC.106.065501}, the no-core shell model (NCSM)
\cite{PhysRevC.103.014315}, In-medium similarity renormalization group (IMSRG) \cite{PhysRevC.102.014302}. 

The $^{136}$Xe is one of the important $0\nu\beta\beta$ decay candidates with an immense experimental interest all over the globe, such as KamLAND-Zen \cite{PhysRevLett.130.051801}, nEXO \cite{nEXO:2021ujk}, EXO-200 \cite{EXO-200:2019rkq}, PandaX-III \cite{Zhang:2023ywy}, and NEXT \cite{NEXT:2023daz}. This serves as motivation for us to enhance the reliability of NMEs regarding the 0$\nu\beta\beta$ decay of $^{136}$Xe, employing the nuclear shell model. 

The $0\nu\beta\beta$ decay of $^{136}$Xe occurs as 
\begin{equation}
    ^{136}\text{Xe}\rightarrow^{136}\text{Ba}+e^-+e^-.
\end{equation}
Previously, the NMEs for the light neutrino-exchange mechanism of 0$\nu\beta\beta$ decay for $^{136}$Xe was calculated using closure approximation in the nuclear shell model in Refs. \cite{Menendez:2008jp,PhysRevLett.110.222502,Neacsu:2016njp,Menendez:2017fdf,PhysRevC.98.064324,PhysRevC.98.035502,PhysRevC.101.035504,PhysRevLett.100.052503,Coraggio:2020iht,PhysRevC.101.044315,Higashiyama:2020ohf,Jokiniemi:2021qqv,PhysRevC.107.045501}. This approach approximates the effects of excitation energy of all virtual intermediate states of the $0\nu\beta\beta$ decay with a single, constant energy value. However, this simplification neglects the contributions of actual individual excited states. The nonclosure approach offers a more comprehensive calculation by incorporating the actual excitation energies for each spin-parity of these intermediate states. The nonclosure approach can improve the reliability of NMEs for $0\nu\beta\beta$ decay and avoid the problem of picking the correct closure energy value. The nonclosure approach has gained popularity in recent years due to increasing computational resources. This approach was applied in the nuclear shell model calculations of $0\nu\beta\beta$ decay for $^{48}$Ca \cite{PhysRevC.88.064312,PhysRevC.101.014307,PhysRevC.102.034317} $^{76}$Ge \cite{PhysRevC.93.044334,PhysRevC.90.051301}, $^{82}$Se \cite{PhysRevC.89.054304}, $^{124}$Sn \cite{PhysRevC.109.024301}, and briefly for $^{136}$Xe \cite{PhysRevC.91.024309}.

The present work aims to conduct a detailed study of $0\nu\beta\beta$ decay in $^{136}$Xe using the nonclosure approach. This approach is employed to calculate the reliable NMEs within the nuclear shell model. The study examines the effects of the excitation energy of numerous intermediate states and various nuclear structure aspects on the NMEs for the $0\nu\beta\beta$ decay of $^{136}$Xe. 

The paper is organized as follows. Section \ref{sec:II} outlines the theoretical formalism for computing the NMEs in the nuclear shell model for $0\nu\beta\beta$ decay with closure and nonclosure approaches. Section \ref{sec:III} 
explores our findings: calculations, comparisons to prior research, and discussion. Finally, in Section \ref{sec:IV}, we summarize the main conclusions of our work.

\section{\label{sec:II}Theoretical Formalism}

For the light neutrino-exchange mechanism of $0\nu\beta\beta$ decay, the inverse of half-life is \cite{PhysRevC.60.055502}
\begin{equation}
    [T^{0\nu}_\frac{1}{2}]^{-1}=G^{0\nu}g_{A}^{4}|M^{0\nu}|^2\left(\frac{\langle m_{\beta\beta}\rangle}{m_e}\right)^2,
\end{equation}
where $G^{0\nu}$ is phase-space factor \cite{PhysRevC.85.034316}, $M^{0\nu}$ is the total NME for the light neutrino-exchange mechanism, and the effective Majorana mass $\langle m_{\beta\beta}\rangle$ is defined by the neutrino mass eigenvalues $m_k$ and the neutrino mixing matrix elements $U_{ek}$, given in Eq. (3) of Ref. \cite{PhysRevC.101.014307}.

The total nuclear matrix element $M^{0\nu}$ contains Gamow-Teller ($M_{GT}$), Fermi ($M_{F}$), and tensor ($M_{T}$) matrix elements  as given by \cite{PhysRevC.60.055502}
\begin{equation}
M^{0\nu}=M_{GT}-\left(\frac{g_V}{g_A}\right)^{2}M_{F}+M_{T},
\end{equation}
where $g_V$ and $g_A$ are the vector and axial-vector constants, respectively. In the present work, $g_V$=1 and the bare value of $g_A$=1.27 are used. The matrix elements $M_{GT}$, $M_{F}$, and $M_{T}$ of the two-body transition operator $O_{12}^\alpha$ of $0\nu\beta\beta$ decay can be expressed as \cite{PhysRevLett.113.262501}:
\begin{eqnarray}
\label{Eq:NMEMAIN}
&&{M}_{\alpha}=\langle f|f_{Jastrow}(r)O_{12}^\alpha f_{Jastrow}(r)|i\rangle,
\end{eqnarray}
where $\alpha\in{F, GT, T}$, and in the present case, $|i\rangle$ corresponds to the $0^+$ ground state of the parent nucleus $^{136}$Xe, and $|f\rangle$ corresponds to the $0^+$ ground state of the granddaughter nucleus $^{136}$Ba. The $O_{12}^{\alpha}$ is the scalar two-particle transition operator of $0\nu\beta\beta$ decay that incorporates both spin and radial neutrino potential operators. In standard Jastrow approach, the $f_{Jastrow}(r)$ includes the effects of SRC \cite{PhysRevC.79.055501,vogel2012nuclear} which is defined as 
\begin{equation}
\label{eq:src}
  f_{Jastrow}(r)=1-ce^{-ar^{2}}(1-br^{2}).  
\end{equation}
In the literature, three different short-range correlation (SRC) parameterizations are used in the Jastrow approach: Miller-Spencer, Charge-Dependent Bonn (CD-Bonn), and Argonne V18 (AV18) \cite{PhysRevC.81.024321}. The parameters $a$, $b$, and $c$ for these SRC parameterizations are detailed in Ref. \cite{PhysRevC.81.024321}. The Jastrow-like function approach to incorporating SRC effects is extensively utilized in Refs. \cite{PhysRevC.81.024321,Menendez:2008jp,Sarkar:2023vdl}. Recently, the authors of Refs. \cite{kortelainen2007short,PhysRevC.75.051303} have proposed an alternative method called the Unitary Correlation Operator Method (UCOM) for estimating SRC effects. This study focuses exclusively on the Jastrow-type approach for estimating SRC effects. Detailed descriptions of incorporating SRC effects using different approaches can be found in Refs. \cite{vogel2012nuclear,PhysRevC.79.055501}.


The two-body transition operators $O_{12}^{\alpha}$ for $0\nu\beta\beta$ decay in light neutrino-exchange mechanism are can be written as \cite{PhysRevC.88.064312}
\begin{eqnarray}
\label{eq:ncoperator}
O_{12}^{GT}&&=\tau_{1-}\tau_{2-}(\mathbf{\sigma_1.\sigma_2)}H_{GT}(r,E_k),
\nonumber\\
O_{12}^{F}&&=\tau_{1-}\tau_{2-}H_{F}(r,E_k),
\\
O_{12}^{T}&&=\tau_{1-}\tau_{2-}S_{12}H_{T}(r,E_k),
\nonumber
\end{eqnarray}
where $\tau$ is the isospin annihilation operator, $\mathbf{r=r_1-r_2}$ is the inter-nucleon distance of the decaying nucleons, and the tensor spin operator $S_{12}$ is defined as $S_{12}=3(\mathbf{\sigma_1 .\hat{r})(\sigma_2.\hat{r})-(\sigma_1.\sigma_2)}$.

The radial neutrino potential with explicit dependence on the energy of the intermediate states is given by \cite{PhysRevC.88.064312}

\begin{equation}
\label{eq:npnc}
H_\alpha (r,E_{k})=\frac{2R}{\pi}\int_{0}^{\infty}\frac{f_\alpha(q,r)qdq}{q+E_{k}-(E_{i}+E_{f})/2},
\end{equation}
where $R$ is the radius of the parent nucleus, $q$ is the momentum of the virtual Majorana neutrino, and $E_{i}$, $E_{k}$, and $E_{f}$ are the energies of the initial, intermediate, and final states, respectively. The term $f_\alpha(q,r) = j_{p}(q,r)h_\alpha(q^2)$, where $j_{p}(q,r)$ is the spherical Bessel function with $p=0$ for Fermi and Gamow-Teller, and $p=2$ for tensor NMEs. The term $h_\alpha(q^2)$ accounts for the effects of finite nucleon size (FNS) and higher-order currents (HOC) as described in Refs. \cite{PhysRevC.60.055502,PhysRevC.79.055501}. In the expression for $h_\alpha(q^2)$, the parameters $M_V$ and $M_A$ are 850 MeV and 1086 MeV, respectively, and $\mu_p - \mu_n=$4.7 is used in the calculations \cite{PhysRevC.81.024321}.

Two different approaches to evaluating neutrino potential integral in NME calculations are closure and nonclosure approaches.
In closure approach, one approximates the term $E_{k} - (E_{i}+E_{f})/2$ in the denominator of the neutrino potential of Eq. (\ref{eq:npnc}) with a constant closure energy ($\langle E\rangle$) such that $E_{k}-(E_{i}+E_{f})/2\rightarrow \langle E\rangle$ \cite{PhysRevC.81.024321}. The closure approximation simplifies calculations by eliminating the need to account for a large number of allowed spin-parity states of intermediate nuclei, which can be computationally challenging for higher-mass isotopes such as $^{136}$Xe in the nuclear shell model. However, the accuracy of this method relies on the selection of an appropriate closure energy.

In the nonclosure approach, the neutrino potential of Eq.(\ref{eq:npnc}) is computed explicitly in NMEs calculation by considering energy $E_k$ of a large number of states $|k\rangle$ of the virtual intermediate nucleus ($^{136}$Cs for the present case). 
The term $E_{k} - (E_{i}+E_{f})/2$ in the denominator of the neutrino potential in Eq. (\ref{eq:npnc}) is expressed as a function of the excitation energy ($E_k^*$) of the intermediate state ($|k\rangle$) as $E_{k} - (E_i+E_f)/2 \rightarrow Q_{\beta \beta}(0^+)/2 + \Delta M + E_k^*$ \cite{PhysRevC.88.064312}. Here, $Q_{\beta \beta}(0^+)$ is the Q value for the $0\nu\beta \beta$ decay of $^{136}$Xe, $\Delta M$ is the mass difference between the $^{136}$Cs and $^{136}$Xe isotopes, and $E_k^*$ is the excitation energy of the intermediate states $|k\rangle$ with different allowed spin-parities of $^{136}$Cs.

This paper primarily focuses on employing the nonclosure approach to account for the realistic effects of at least two hundred states for each spin-parity of the virtual intermediate nucleus $^{136}$Cs. Additionally, it derives the optimal closure energy required to reproduce the NMEs obtained from the nonclosure approach using the closure method.

\begin{table*}
\caption{\label{tab:nmet1} The nuclear matrix elements $M_F$, $M_{GT}$, $M_T$, and $M^{0\nu}$ for the $0\nu\beta\beta$ decay of $^{136}$Xe (light neutrino-exchange mechanism) are calculated using the GCN5082 interaction with the running nonclosure method across different SRC parameterizations. The SRC type labeled "None" includes only the effects of FNS and HOC.}
\begin{ruledtabular}
\begin{tabular}{ccc}
NME Type&SRC Type&Nonclosure NME\\ \hline
$M_F$&None
&-0.458

\\
$M_F$&Miller-Spencer
&-0.322

\\
$M_F$&CD-Bonn
&-0.489

\\
$M_F$&AV18
&-0.451

\\
\\
$M_{GT}$&None
    &1.693

\\
$M_{GT}$&Miller-Spencer
&1.225

\\
$M_{GT}$&CD-Bonn
&1.743

\\
$M_{GT}$&AV18
&1.611

\\
\\
$M_T$&None
    &0.013

\\
$M_T$&Miller-Spencer
&0.012

\\
$M_T$&CD-Bonn
&0.012

\\
$M_T$&AV18
&0.012

\\
\\
$M^{0\nu}$&None
    &1.990

\\
$M^{0\nu}$&Miller-Spencer
&1.437

\\
$M^{0\nu}$&CD-Bonn
&2.058

\\
$M^{0\nu}$&AV18
&1.903

\\
\end{tabular}
\end{ruledtabular}
\end{table*}
\begin{table*}
\caption{\label{tab:nmecomparison} The comparison of total NME ($M^{0\nu}$) for the light neutrino-exchange mechanism of $0\nu\beta\beta$ decay in $^{136}$Xe, calculated using various many-body nuclear models. This comparison includes both nonclosure and closure approaches with different SRC parameterizations and $g_{A}$.
}
\begin{ruledtabular}
\begin{tabular}{cccccc}
Nuclear Model&Reference&Approximation&$g_{A}$&SRC Type& Total NME ($M^{0\nu}$)\\ \hline
ISM&Current Study&Nonclosure&1.270&CD-Bonn&2.06\\

ISM&Ref. \cite{PhysRevC.98.035502}&Closure&1.270&CD-Bonn&1.74
\\

ISM&Ref. \cite{PhysRevC.91.024309}&Closure&1.254&CD-Bonn&1.76
\\

ISM&Ref. \cite{Menendez:2008jp}&Closure&1.250&UCOM& 2.19
\\
QRPA&Ref. \cite{PhysRevC.91.024613}&Closure&1.260&CD-Bonn& 2.91
\\
QRPA&Ref. \cite{PhysRevD.90.096010}&Closure&1.269&CD-Bonn& 2.46
\\
QRPA&Ref. \cite{PhysRevD.90.096010}&Closure&1.269&AV18& 2.18
\\
GCM&Ref. \cite{PhysRevC.98.064324}&Closure&1.254&CD-Bonn& 2.35
\\
IBM2&Ref. \cite{PhysRevC.91.034304}&Closure&1.269&AV18& 3.05
\\
EDF&Ref. \cite{PhysRevLett.105.252503}&Closure&1.250&UCOM& 4.20
\\
REDF&Ref. \cite{PhysRevC.91.024316}&Closure&1.254&None& 4.32
\\

\end{tabular}
\end{ruledtabular}
\end{table*}

The method based on the nonclosure approach is referred to as the running nonclosure method \cite{PhysRevC.88.064312}, due to the limitation of calculating only a finite number of intermediate states with current computational resources. The partial NMEs for the transition operator of Eq. (\ref{eq:ncoperator}) in the running nonclosure method is defined as \cite{PhysRevC.102.034317}
\begin{widetext}
\begin{eqnarray}
\label{eq:rncpartial}
&&M_{\alpha}(J_{k},J,E_{k}^{*})=\sum_{k'_{1}k'_{2}k_{1}k_{2}}\sqrt{(2J_{k}+1)(2J_{k}+1)(2J+1)}\times(-1)^{j_{k1}+j_{k2}+J}
\left\{ \begin{array}{ccc}
j_{k1^{'}} & j_{k1} & J_{k}\\
j_{k2} & j_{k2^{'}} & J
\end{array}\right\}\nonumber\\ &&\times\text{OBTD}(k,f,k'_{2},k_{2},J_{k})\times \text{OBTD}(k,i,k'_{1},k_{1},J_{k})\langle k_1',k_2':J||f_{Jastrow}(r)O_{12}^\alpha f_{Jastrow}(r)||k_1,k_2:J\rangle.\nonumber\\
\label{eq:mjjkrc}
\end{eqnarray}
\end{widetext}
Here, $k_1$ represents a set of quantum numbers $(n_1, l_1, j_1)$ corresponding to an orbit, with similar notations for $k_2$, $k_1'$, and $k_2'$. In this study, $k_1$ (and others) corresponds to the quantum numbers associated with the $0g_{7/2}$, $1d_{5/2}$, $1d_{3/2}$, $2s_{1/2}$, and $0h_{11/2}$ orbits within the jj55 model space. $J$ denotes the allowed spin-parity of the two decaying neutrons and the created protons, while $J_k$ represents the allowed spin-parity of the intermediate states $|k\rangle$.
The complete expression for the non-antisymmetric reduced two-body matrix elements ($\langle k_1', k_2': J || f_{Jastrow}(r){O}_{12}^\alpha f_{Jastrow}(r) || k_1, k_2: J \rangle$) used in the running nonclosure method is detailed in Ref. \cite{PhysRevC.88.064312}. The one-body transition density (OBTD) represents the matrix elements of neutron annihilation and proton creation operators. These are expressed within the proton-neutron formalism, as shown in Eq. (41) of Ref. \cite{PhysRevC.102.034317}. When one approximates $E_{k}-(E_{i}+E_{f})/2$ as $\langle E\rangle$ in Eq. (\ref{eq:rncpartial}), it yields the expression of the NME in the running closure method, which is also employed in this study. 

Finally, in the running nonclosure method, NMEs are computed by summing over all intermediate states $|k\rangle$, with excitation energies $E_k^*$ up to a specified cutoff value $E_c$, given by \cite{PhysRevC.88.064312}
\begin{eqnarray}
{M}_{\alpha}(E_c)=\sum_{J_k,J,E_{k}^{*}\leqslant E_c}{M}_{\alpha}(J_{k},J,E_{k}^{*}). 
\end{eqnarray}
The NMEs demonstrate convergence when the $E_c$ values are large enough to encompass all relevant intermediate states.

\section{\label{sec:III}Results and Discussions}
\begin{figure*}
\includegraphics[trim=0cm 0.5cm 0cm 0.5cm,width=\linewidth]{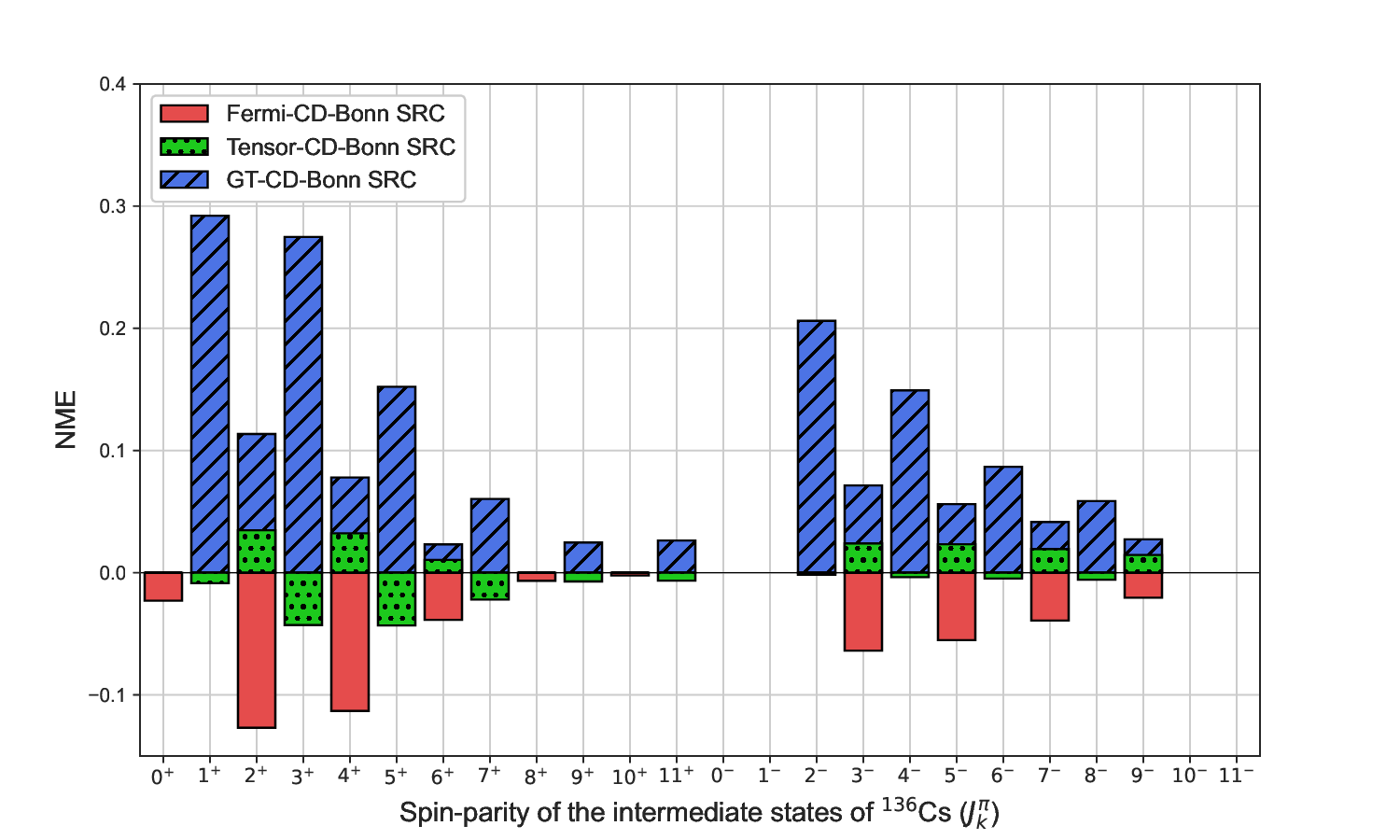}

\caption{\label{fig:NMEvsJk}The figure illustrates the contributions of various spin-parity states $J_k^{\pi}$ of the virtual intermediate nucleus $^{136}$Cs to the NMEs for the light neutrino-exchange mechanism in the $0\nu\beta\beta$ decay of $^{136}$Xe. These NMEs are computed using the running nonclosure method with the GCN5082 effective interaction for CD-Bonn SRC parametrization.}
\end{figure*}
In this study of the $0\nu\beta\beta$ decay of $^{136}$Xe, we utilized the KSHELL code \cite{Shimizu:2019xcd} to diagonalize the shell model Hamiltonian, allowing us to determine the wave functions and energies for the initial ($^{136}$Xe), intermediate ($^{136}$Cs), and final ($^{136}$Ba) nuclei involved in the decay process. The calculations employed the GCN5082 shell model Hamiltonian \cite{PhysRevC.82.064304} within the jj55 model space, also used in previous research \cite{Menendez:2008jp}. While the SVD Hamiltonian has been applied in the jj55 model space in earlier study \cite{PhysRevC.98.035502}, it was not available for this investigation. 

For the intermediate nucleus $^{136}$Cs involved in the $0\nu\beta\beta$ decay of $^{136}$Xe, we computed the lowest 200 energy states for each permissible spin-parity $J_k^{\pi}$. These calculated wave functions were used for generating an extensive set of OBTDs. Finally, we developed programming to compute the non-antisymmetric reduced two-body matrix elements (TBMEs) essential for running nonclosure and closure methods of NME calculations. 
\begin{figure*}
\centering
\includegraphics[trim=1cm 1.5cm 1.5cm 1cm,width=\linewidth]{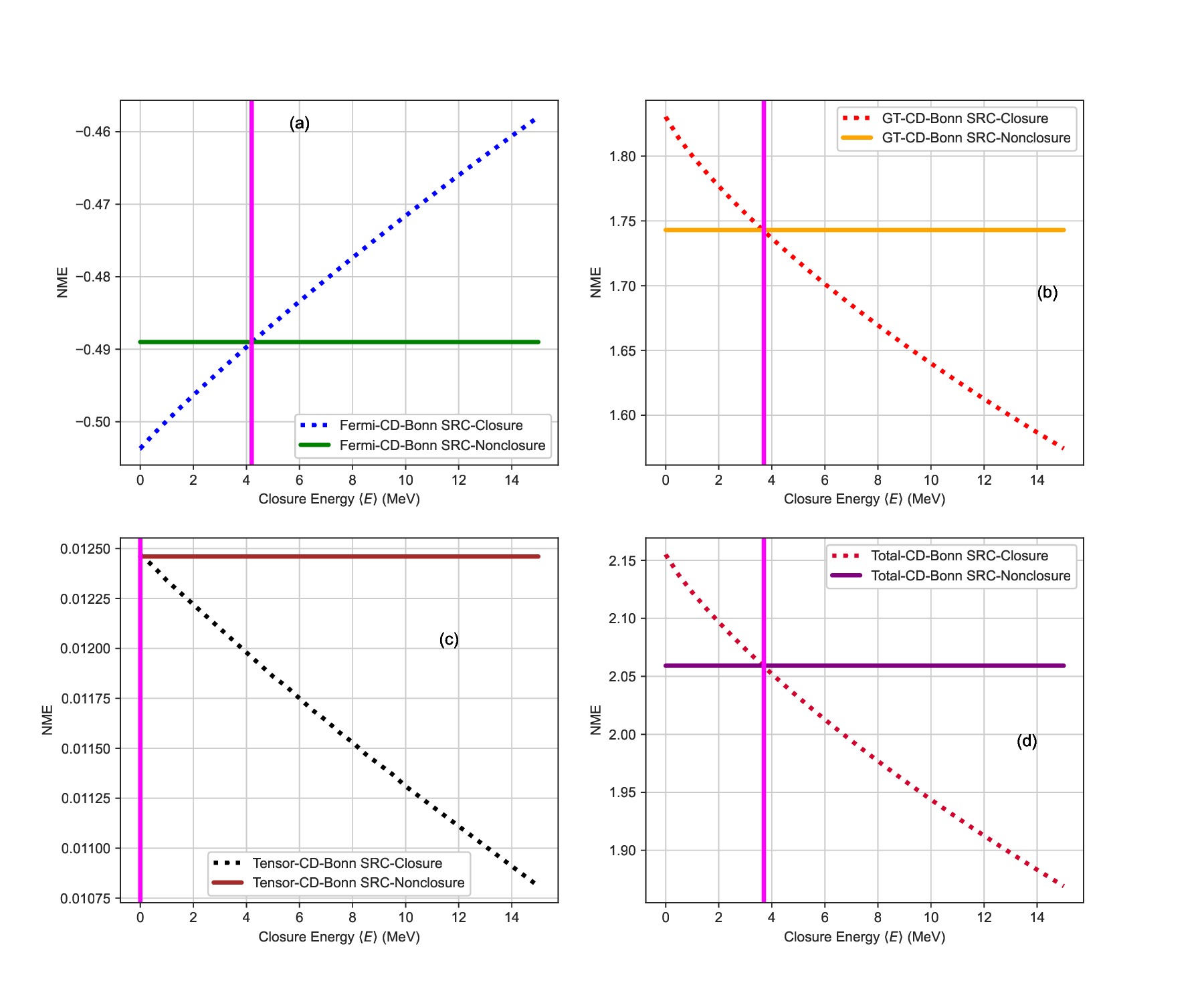}
\caption{\label{fig:optimalclosure}
The figure demonstrates the closure NME dependencies for (a) Fermi, (b) Gamow-Teller, (c) Tensor, and (d) Total types as functions of closure energy $\langle E\rangle$ using CD-Bonn SRC parametrizations for the $0\nu\beta\beta$ decay of $^{136}$Xe. It highlights where closure and nonclosure NMEs intersect, marking the optimal closure energy where these NMEs overlap, indicated by a vertical magenta line. The closure NMEs are shown as dotted lines (various colors), while nonclosure NMEs are depicted as solid lines (various colors). Notably, the optimal closure energy for which closure and nonclosure NMEs converge is consistent across different SRC parametrizations.}
\end{figure*}

\begin{figure*}
\centering
\includegraphics[trim=1.5cm 1cm 2.5cm 1cm,width=\linewidth]{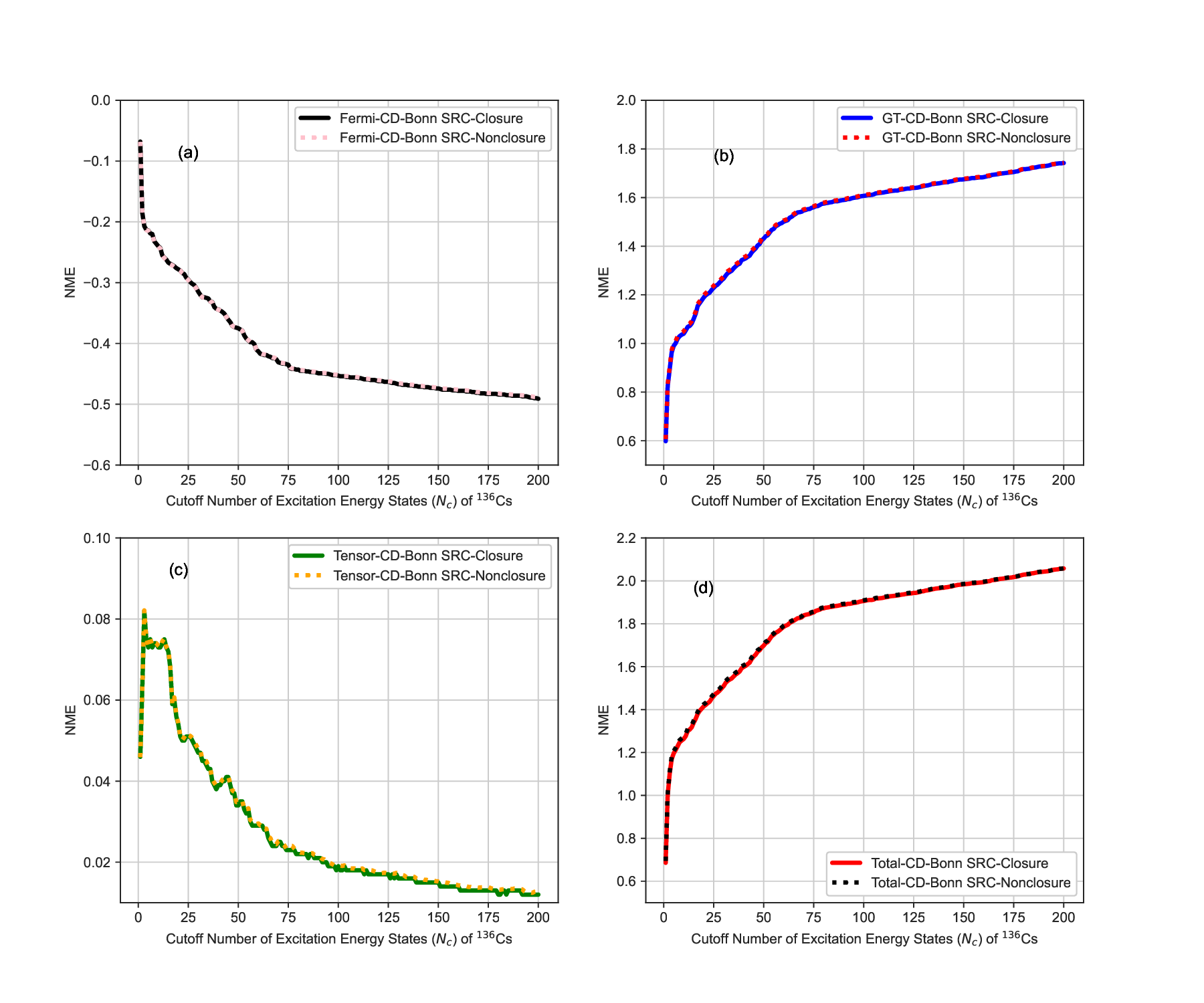}
\caption{\label{fig:nmevsnk}Variation of (a) Fermi, (b) Gamow-Teller, (c) tensor, and (d) total NMEs for the $0\nu\beta\beta$ decay of $^{136}$Xe (the light neutrino-exchange mechanism) with the cutoff number of excitation energy states ($N_c$) of the virtual intermediate nucleus $^{136}$Cs. The NMEs are computed using the GCN5082 interaction for the CD-Bonn SRC parametrization in both the running nonclosure method (dotted lines) and the closure method (solid lines). The closure method employs an optimal closure energy of $\langle E\rangle \approx 3.7$ MeV for all types of NME.}
\end{figure*}

Table \ref{tab:nmet1} provides a detailed summary of the calculated NMEs for the $0\nu\beta\beta$ decay of $^{136}$Xe, using the nonclosure method within the nuclear shell model framework. All NMEs are computed with corrections for FNS and HOC. The table also includes NMEs derived from three different short-range correlation (SRC) parameterizations: Miller-Spencer, CD-Bonn, and AV18. The results indicate that Gamow-Teller (GT) NMEs are dominant compared to the Fermi and tensor NMEs, underscoring their critical role in the decay process. Significant variations in NMEs are observed depending on the type of SRC employed, with the Miller-Spencer SRC exhibiting the most pronounced impact. For comparison, the "None" SRC type accounts for only FNS and HOC effects.

In Table \ref{tab:nmecomparison}, we present the newly calculated total NME ($M^{0\nu}$) for the $0\nu\beta\beta$ decay of $^{136}$Xe. This calculation employs the interacting shell model in the nonclosure approximation with the CD-Bonn SRC for the light neutrino-exchange mechanism. Additionally, we compare the results with NMEs reported from various many-body nuclear models, SRCs, and approximations. The NMEs across different models and approximations range from 1.74 to 4.32.

In Ref. \cite{PhysRevC.98.035502}, the $M^{0\nu}$ for the CD-Bonn SRC was calculated to be 1.74 using the shell model with a closure energy of 3.5 MeV. On the other hand, Ref. \cite{Menendez:2008jp} reports an NME of 2.19, calculated using the shell model with UCOM SRC in the closure approximation. In our present study, the newly calculated $M^{0\nu}$ is 2.06, determined using the shell model in a nonclosure approach with CD-Bonn SRC. This value is about 18\% larger than the results in Ref. \cite{PhysRevC.98.035502} and about 6\% smaller than those in Ref. \cite{Menendez:2008jp}. These differences may arise from the Hamiltonian, SRC type, and closure energy used in the earlier studies.

It is observed that NMEs calculated with nuclear models other than the shell model tend to be larger. This significant variation in NMEs across different models highlights an ongoing challenge to reconcile these differences and achieve more consistent results in the field.

\subsection{Contributions of spin-parity ($J_{k}^{\pi}$) of $^{136}$Cs on $0\nu\beta\beta$ decay NMEs for $^{136}$Xe}
To understand the impact of nuclear structure on $0\nu\beta\beta$ decay of $^{136}$Xe, we first analyze the contributions from specific spin-parity of the intermediate nucleus $^{136}$Cs. Using Eq. (\ref{eq:rncpartial}), we compute the partial NMEs for each spin-parity as follows:

\begin{eqnarray}
{M}_{\alpha}(E_c,J_k)=\sum_{J,E_{k}^{*}\leqslant E_c}{M}_{\alpha}(J_{k},J,E_{k}^{*}),
\end{eqnarray}
where $J_k^{\pi}$ denotes the spin-parity state of $^{136}$Cs.

Figure~\ref{fig:NMEvsJk} shows the contributions of each $J_k^{\pi}$ of $^{136}$Cs on the Fermi, Gamow-Teller, and tensor NMEs, calculated using the running nonclosure method with CD-Bonn SRC.

In Fermi-type NMEs, all contributions from various $J_k^{\pi}$ states are negative. Conversely, Gamow-Teller NMEs show consistently positive contributions. Tensor-type NMEs, however, feature both positive and negative contributions depending on the $J_k^{\pi}$ state.

The 2$^{+}$ state contributes the most significantly to Fermi-type NMEs, with substantial contributions also from the 4$^{+}$, 6$^{+}$, 3$^{-}$, 5$^{-}$, and 7$^{-}$ states.

For Gamow-Teller NMEs, the 1$^{+}$ state is the primary contributor, while significant contributions also come from the 3$^{+}$, 5$^{+}$, 2$^{-}$, and 4$^{-}$ states.

In tensor-type NMEs, positive contributions are most prominent from the 2$^{+}$, 4$^{+}$, 3$^{-}$, and 5$^{-}$ states, while negative contributions are notable from the 3$^{+}$, 5$^{+}$, and 7$^{+}$ states.

This pattern of NME variation with $J_k^{\pi}$ persists across different SRC parameterizations.
\subsection{The optimal closure energy for light neutrino-exchange $0\nu\beta\beta$ decay of $^{136}$Xe}
At the optimal closure energy, the NMEs calculated using the running closure method align with those obtained from the running nonclosure method. This congruence is advantageous because the closure approach is computationally less demanding, eliminating the need to evaluate the neutrino potential integral for each intermediate state energy.

To determine the optimal closure energy for the $0\nu\beta\beta$ decay of $^{136}$Xe in the light neutrino-exchange mechanism, we examined the variations of Fermi, Gamow-Teller, tensor, and total NMEs with different closure energies, as depicted in Fig. \ref{fig:optimalclosure}. This analysis includes both closure and nonclosure NMEs, with the intersection point indicating the optimal closure energy, marked by a vertical magenta line. The pattern observed for the CD-Bonn SRC is consistent across other SRC types, showing similar optimal closure energies.

The results indicate that the optimal closure energy for the Fermi-type NME with CD-Bonn SRC is approximately 4.1 MeV. For GT NMEs, the optimal closure energy is around 3.7 MeV, and for tensor-type NMEs, it is near 0 MeV. The total NMEs are predominantly influenced by the GT component, leading to an optimal closure energy of approximately 3.7 MeV for the total NMEs.

In the next subsection, for simplicity and consistency, we use 3.7 MeV (derived from the total NME) as the optimal closure energy for all NME types in the $0\nu\beta\beta$ decay of $^{136}$Xe via the light neutrino-exchange mechanism. This choice demonstrates that the NMEs calculated using the closure approach closely reproduce those obtained from the nonclosure approach across the number of spin-parity states considered for $^{136}$Cs for different types of NME. 
\subsection{Convergence of NMEs for $0\nu\beta\beta$ of $^{136}$Xe with the cutoff number of states ($N_c$) of $^{136}$Cs}
In the concluding part of this study, we evaluate the convergence behavior of NMEs for the $0\nu\beta\beta$ decay of $^{136}$Xe with respect to the cutoff number of energy states ($N_c$) considered for each spin-parity ($J_k^{\pi}$) of the intermediate nucleus $^{136}$Cs. We express the NMEs as a function of $N_c$ using the running nonclosure method:

\begin{eqnarray}
{M}_{\alpha}(N_c)=\sum_{J_k,J,N_k\leqslant N_c} {M}_{\alpha}(J_{k},J,N_k),
\end{eqnarray}

where ${M}_{\alpha}(J_{k}, J, N_k)$ follows the definition provided in Eq. (\ref{eq:rncpartial}). The running closure method NME as a function $N_c$ is the same as above, except the energy term of the intermediate nucleus is approximated with constant closure energy, as discussed in the formalism part.

Figure~\ref{fig:nmevsnk} shows the dependence of Fermi, Gamow-Teller, tensor, and total NMEs on $N_c$ for $^{136}$Cs in both the running closure and nonclosure methods. The figure focuses on the CD-Bonn SRC parametrization, though similar trends are observed for other SRCs. For the closure method, we use an optimal closure energy of 3.7 MeV for all types of NME.

In this study, the maximum number of intermediate states considered for each allowed $J_k^{\pi}$ of $^{136}$Cs is $N_c=200$, limited by current computational capabilities. We find that even with 200 intermediate states per spin-parity, the NMEs have not fully converged in either the running closure or nonclosure methods. The tensor-type NMEs, while less saturated, have minimal impact compared to the dominant Gamow-Teller and Fermi-type NMEs.

Expanding our computational efforts beyond $N_c=200$ proved challenging for now, especially for the heavier isotope $^{136}$Xe. However, we believe that calculated NMEs are within reasonable accuracy. Furthermore, our analysis revealed the crossover of running closure and nonclosure NMEs at an optimal closure energy of approximately 3.7 MeV. This underscores the dominance of lower-energy states, with the major contributions adequately captured within $N_c=200$ for each $J_k^{\pi}$ configuration of $^{136}$Cs.

To illustrate the efficacy of the determined optimal closure energy, Fig.~\ref{fig:nmevsnk} demonstrates that the running closure method, employing a closure energy of 3.7 MeV for all types of NME, accurately reproduces the NMEs obtained through the running nonclosure method across all considered cutoff numbers of states, up to $N_c=200$. This suggests that the trend may persist beyond $N_c=200$. Therefore, future endeavors aiming to calculate states of $^{136}$Cs beyond $N_c=200$ can leverage the established optimal energy of 3.7 MeV. By employing the closure method, one can efficiently reproduce nonclosure NMEs beyond $N_c=200$, minimizing computational requirements.
\section{\label{sec:IV}Summary}
The $^{136}$Xe is an important candidate for global experimental studies of $0\nu\beta\beta$ decay. 

In this study, we calculated the NMEs for $0\nu\beta\beta$ decay of $^{136}$Xe in the light neutrino-exchange mechanism within the nuclear shell model framework. The calculations employed mainly the nonclosure approach, explicitly considering excitation energies for 200 states of each spin-parity in the intermediate nucleus $^{136}$Cs, which enhances the accuracy of the NMEs.

Our findings indicate up to 18\% variation in NMEs compared to recent calculations using the closure approach with different Hamiltonians, SRC, and closure energies in the nuclear shell model. This discrepancy may be attributed to variations in the Hamiltonian, SRC, or closure energy choices in previous studies.

We also analyzed the NMEs' dependence on the spin-parity of the intermediate states in $^{136}$Cs. For Gamow-Teller NMEs, contributions from each spin-parity state were positive, while for Fermi NMEs, the contributions were negative.

An optimal closure energy of approximately 3.7 MeV was identified, where the closure NMEs align with the nonclosure NMEs for the $0\nu\beta\beta$ decay of $^{136}$Xe (light neutrino-exchange mechanism). This optimal closure energy can be utilized in future closure approach calculations, significantly reducing computational demands.

The convergence of NMEs with respect to the number of intermediate states was also investigated. Although full convergence was not achieved with 200 intermediate states for each spin-parity of $^{136}$Cs, expanding beyond this number is currently computationally challenging for us. However, we identified the crossover point between running nonclosure and running closure NMEs at the optimal closure energy of 3.7 MeV, effectively reproducing the nonclosure NMEs using the closure approach for all considered cutoff numbers of states for each spin-parity of the intermediate nucleus $^{136}$Cs.

Future studies can extend this nonclosure approach to investigate other non-standard mechanisms of $0\nu\beta\beta$ decay, potentially broadening our understanding of the process.
\begin{acknowledgments}
S.S. thanks the Anusandhan National Research Foundation (ANRF), Government of India, for the Science and Engineering Research Board (SERB) National Postdoctoral Fellowship under grant no. PDF/2022/003729. 

S.S. and R.C. acknowledge the National Supercomputing Mission (NSM) for providing computing resources of ‘PARAM Ganga’ at the Indian Institute of Technology Roorkee, which is implemented by C-DAC and supported by the Ministry of Electronics and Information Technology (MeitY) and Department of Science and Technology (DST), Government of India.
\end{acknowledgments}


\bibliography{main}

\end{document}